\documentstyle[epsf,twocolumn,prd,aps,floats]{revtex}
\begin{document}

\draft

%
%
\newcommand{\nc}{\newcommand}
\nc{\bea}{\begin{eqnarray}}
\nc{\eea}{\end{eqnarray}}
\nc{\beq}{\begin{equation}}
\nc{\eeq}{\end{equation}}
\nc{\bi}{\begin{itemize}}
\nc{\ei}{\end{itemize}}
\nc{\la}[1]{\label{#1}}

\nc{\D}{{\rm D}}
\nc{\EH}{{}^3{\rm H}}
\nc{\EHe}{{}^3{\rm He}}
\nc{\UHe}{{}^4{\rm He}}
\nc{\GLi}{{}^6{\rm Li}}
\nc{\ZLi}{{}^7{\rm Li}}
\nc{\ZBe}{{}^7{\rm Be}}
\nc{\DH}{{\rm D}/{\rm H}}
\nc{\EHeH}{^3{\rm He}/{\rm H}}
\nc{\ZLiH}{{}^7{\rm Li}/{\rm H}}

\nc{\et}{\eta_{10}}
\nc{\GeV}{\mbox{ GeV}}
\nc{\MeV}{\mbox{ MeV}}
\nc{\keV}{\mbox{ keV}}
\nc{\eV}{\mbox{ eV}}

\nc{\Neff}{N_{\rm eff}}
\nc{\DNeff}{\Delta N_{\rm eff}}
\nc{\gstar}{g_\ast}

%
%
\nc{\etal}{{\it et al.}}
\nc{\ibid}[3]{{\it ibid.\ }{{\bf #1}{, #2}{ (#3)}}}
\nc{\x}[1]{}

\nc{\AJ}[3]{{Astron.~J.\ }{{\bf #1}{, #2}{ (#3)}}}
\nc{\anap}[3]{{Astron.\ Astrophys.\ }{{\bf #1}{, #2}{ (#3)}}}
\nc{\ApJ}[3]{{Astrophys.~J.\ }{{\bf #1}{, #2}{ (#3)}}}
\nc{\apjl}[3]{{Astrophys.~J.\ Lett.\ }{{\bf #1}{, #2}{ (#3)}}}
\nc{\app}[3]{{Astropart.\ Phys.\ }{{\bf #1}{, #2}{ (#3)}}}
\nc{\araa}[3]{{Ann.\ Rev.\ Astron.\ Astrophys.\ }{{\bf #1}{, #2}{ (#3)}}}
\nc{\arns}[3]{{Ann.\ Rev.\ Nucl.\ Sci.\ }{{\bf #1}{, #2}{ (#3)}}}
\nc{\arnps}[3]{{Ann.\ Rev.\ Nucl.\ and Part.\ Sci.\ }{{\bf #1}{, #2}{ (#3)}}}
\nc{\epj}[3]{{Eur.\ Phys.\ J.\ }{{\bf #1}{, #2}{ (#3)}}}
\nc{\MNRAS}[3]{{Mon.\ Not.\ R.\ Astron.\ Soc.\ }{{\bf #1}{, #2}{ (#3)}}}
\nc{\mpl}[3]{{Mod.\ Phys.\ Lett.\ }{{\bf #1}{, #2}{ (#3)}}}
\nc{\Nat}[3]{{Nature }{{\bf #1}{, #2}{ (#3)}}}
\nc{\ncim}[3]{{Nuov.\ Cim.\ }{{\bf #1}{, #2}{ (#3)}}}
\nc{\nast}[3]{{New Astronomy }{{\bf #1}{, #2}{ (#3)}}}
\nc{\np}[3]{{Nucl.\ Phys.\ }{{\bf #1}{, #2}{ (#3)}}}
\nc{\pr}[3]{{Phys.\ Rev.\ }{{\bf #1}{, #2}{ (#3)}}}
\nc{\PRC}[3]{{Phys.\ Rev.\ C\ }{{\bf #1}{, #2}{ (#3)}}}
\nc{\PRD}[3]{{Phys.\ Rev.\ D\ }{{\bf #1}{, #2}{ (#3)}}}
\nc{\PRL}[3]{{Phys.\ Rev.\ Lett.\ }{{\bf #1}{, #2}{ (#3)}}}
\nc{\PL}[3]{{Phys.\ Lett.\ }{{\bf #1}{, #2}{ (#3)}}}
\nc{\prep}[3]{{Phys.\ Rep.\ }{{\bf #1}{, #2}{ (#3)}}}
\nc{\RMP}[3]{{Rev.\ Mod.\ Phys.\ }{{\bf #1}{, #2}{ (#3)}}}
\nc{\rpp}[3]{{Rep.\ Prog.\ Phys.\ }{{\bf #1}{, #2}{ (#3)}}}
\nc{\zphysA}[3]{{Z.\ Phys.\ A }{{\bf #1}{, #2}{ (#3)}}}

\nc{\pla}[3]{{Plasma Phys.\ }{{\bf #1}{, #2}{ (#3)}}}
\nc{\ndt}[3]{{Nuclear Data Tables\ }{{\bf #1}{, #2}{ (#3)}}}

\date{June 12, 2002}

\wideabs{

\title{Big Bang nucleosynthesis, matter--antimatter regions, extra
relativistic species, and relic gravitational waves}

\author{Massimo Giovannini\thanks{mailm}}
\address{Institute of Theoretical Physics, University of Lausanne,
  BSP-1015 Dorigny, Switzerland}
\author{Elina Keih\"anen\thanks{maile} and Hannu
Kurki-Suonio\thanks{mailh}}
\address{Department of Physical Sciences, University of Helsinki,
          P.O.Box 64, FIN-00014 University of Helsinki, Finland}

\maketitle

\begin{abstract}
Provided that matter--antimatter domains are present at the onset of big bang
nucleosynthesis (BBN), the number of allowed additional relativistic species
increases, compared to the standard scenario when matter--antimatter domains
are absent.
The extra relativistic species may take the form of massless
fermions or even massless bosons, like relic gravitons.
  The number of additional degrees of
freedom compatible with BBN depends, in this framework,
  upon the typical scale of the domains and the
antimatter fraction. Since
the presence of matter--antimatter domains allows a reduction of the
neutron to proton ratio prior to the formation
of $\UHe$,  large amounts of radiation-like energy density
are allowed. Our results are compared with
other constraints on the number of
supplementary relativistic degrees of freedom and with
different nonstandard BBN scenarios where the reduction of the
neutron to proton
ratio occurs via a different physical mechanism. The implications
of our considerations for the upper limits on
stochastic gravitational waves backgrounds of cosmological origin
are outlined.
\end{abstract}

}

%
%

\section{Introduction}

The strongest constraint on additional energy density in the universe
with a radiation-like equation of state
is provided by big bang nucleosynthesis (BBN).
The additional energy
density speeds up the expansion and cooling of the universe,
and, consequently,
the typical  time scale of BBN is reduced in comparison with the
standard case.
The additional
radiation-like energy density may be attributed to some extra relativistic
species whose statistics may be either bosonic or fermionic.
Since the supplementary species may be  fermionic,  they have been
customarily parametrized in terms of the effective number of
neutrino species
\begin{equation}
\Neff = 3 + \DNeff,
\end{equation}
where $\DNeff = 0$ corresponds to the standard case
with no extra energy density.
The standard BBN (SBBN) results are in agreement with the observed abundances
for $\Neff = 2$--$4$, giving thus an upper limit $\DNeff \lesssim 1$
\cite{Copi,BurlesNeff,Lisi,Cyburt}.

Extra radiation-like energy  can be also constrained
by other physical considerations not directly related to BBN.
However, the bounds seem to be much weaker
than the ones provided by BBN.
For instance,
the growth of density perturbations in the early universe is affected by the
radiation/matter energy density ratio, which in the standard case falls below
unity some time before recombination.
The data on CMB anisotropy\cite{CMBdata}
has now improved to the point where it can be used
to provide a
constraint\cite{Hannestad,Orito,Esposito,Kneller,Hansen,Bowen}.
Hannestad\cite{Hannestad} obtains an upper bound $\Neff \lesssim 17$
(95\% confidence), or $\DNeff \lesssim 14$.  Using large scale structure (LSS)
data in connection with the CMB data the same author also obtained a
lower limit
$\Neff \gtrsim 1.5$/$2.5$ (depending on the data set used).

The BBN and CMB limits are complementary.  Both measure the increase in the
expansion rate of the universe, but through completely different physical
phenomena.  They also look at different epochs in the history of the universe,
BBN to $T={\cal{O}}(1\MeV)$, CMB to $T={\cal{O}}(1\eV)$.  Thus
attempts to relax
one of these limits are unlikely to affect the other.

The BBN limits can be relaxed in nonstandard BBN (NSBBN)
scenarios (see the extensive reviews\cite{NSBBNMM} and\cite{NSBBNS},
or \cite{HXNatal}).  For example, the $\UHe$ yield can be affected by changing
the electron neutrino spectrum, like in the model with active-sterile neutrino
mixing\cite{mixing}, or by $\nu_e$ degeneracy, which changes the amount of
thermal $\nu_e$.  This latter scenario is called degenerate BBN
(DBBN)\cite{otherDBBN,Kang}.

In the present paper yet a different scenario will be studied, namely
BBN with matter--antimatter domains.
In short the idea is the following. Consider the
situation where matter--antimatter fluctuations\cite{Stecker,antimatter},
are present at the onset of BBN. The fluctuations will
be described in terms of their typical scale and
the relative amount of antimatter.
In previous studies\cite{ABBN,KS} it has been established that
this type of scenario is rather effective in reducing
the neutron to proton ratio prior to the formation of the
$\UHe$ abundance. The physical reason behind this statement
is the different diffusion scale of protons and neutrons
which can travel into antimatter domains annihilating there.
Since the presence of extra relativistic species typically
causes an increase in the neutron to proton ratio, it seems plausible
that in the case of antimatter BBN (ABBN) more relativistic degrees of freedom
can be accommodated than in the case of standard BBN scenarios.

The additional relativistic
species may be also bosonic, for instance gravitons.
This possible interpretation
of the extra energy density at the BBN epoch was originally
 invoked by Schwartsman \cite{schwartsman} since
the production of gravitons is a generic phenomenon in
Friedmann-Robertson-Walker (FRW) universes \cite{grishchuk}.
Relic gravitational waves decouple below the Planck scale and  their
energy density is constrained as part of the other
extra relativistic species. The BBN bound constrains
the maximal fraction of critical energy
density present today in relic gravitons of cosmological origin.
The experimental efforts are converging towards
the possibility of direct detection of stochastically
distributed gravitational waves \cite{schutz,grisrev}.
Various resonant mass detectors are now operating at a typical frequency
of the order of kHz (Allegro \cite{allegro},
Auriga \cite{auriga}, Explorer \cite{explorer},
Nautilus \cite{nautilus}, Niobe \cite{niobe})
and four Michelson interferometers (GEO-600\cite{geo},TAMA-400
\cite{tama}, LIGO \cite{ligo} and VIRGO
\cite{virgo}) will soon be operating in a wide frequency
band from a few Hz up to $10$ kHz.
It is interesting, in the context of ABBN, to elaborate on
the possibility that the additional relativistic
species are  of gravitational origin.

The plan of our paper is the following. In Section II some basic considerations
on BBN with extra relativistic species will be presented.
In Section III the main features of ABBN will be summarized
in light of the possible presence of additional relativistic
degrees of freedom in the scenario. In Section IV the results
of the analysis of the parameter space of ABBN
will be reported. In Section V the findings of ABBN will be
compared with the case of DBBN, while in Section
VI the constraints on the additional relativistic
species obtained in the framework of ABBN will
be studied together with other constraints coming
from different physical considerations.
Section VII deals mainly with the implications of our results for
the case of relic gravitational waves. Finally Section VIII
contains our concluding remarks.

\section{BBN with Extra Energy Density}

\subsection{Preliminaries}
In the radiation-dominated era the energy density of the universe is
\beq
    \rho \equiv \gstar\left(\frac{\pi^2}{30}\right)T^4.
\eeq
This equation defines the effective number $\gstar$ of relativistic degrees of
freedom\cite{KolbTurner}.
An (ultra)relativistic fermion species with
two internal degrees of freedom
and in thermal equilibrium contributes $2\cdot7/8 = 7/4 = 1.75$ to
$\gstar$.  Before
neutrino decoupling the contributing relativistic particles are photons,
electrons, positrons, and $N_\nu = 3$ species of neutrinos, giving
\beq
    \gstar = \frac{11}{2} + \frac{7}{4}N_\nu = 10.75.
\eeq
The neutrinos have decoupled before electron-positron annihilation so that they
do not contribute to the entropy released in the annihilation.
While they are
relativistic, the neutrinos still retain an equilibrium energy
distribution, but
after the annihilation
their temperature is lower, $T_\nu = (4/11)^{1/3}T$.  Thus
\beq
    \gstar = 2 + \frac{7}{4}N_\nu\left(\frac{T_\nu}{T}\right)^4
           = 2 + 0.454N_\nu = 3.36
\eeq
after electron-positron annihilation.

If we now assume that there are some additional relativistic degrees of
freedom, which also have decoupled by the time of electron-positron
annihilation,  or just some additional component $\rho_x$ to energy
density with a
radiation-like equation of state, $p_x = \rho_x/3$, its effect on the energy
density and expansion rate of the universe is the same as that of having
some (perhaps a fractional number of)
additional neutrino species.  Thus its contribution can be represented by
replacing $N_\nu$ with $\Neff = N_\nu + \DNeff$ in the above.  Before
electron-positron annihilation we have $\rho_x = (7/8)\DNeff\rho_\gamma$
and after electron-positron annihilation we have
$\rho_x = 0.227\DNeff\rho_\gamma$.

The present ratio of the CMB ($T = 2.725$~K) energy density to the critical density is
$\Omega_\gamma \equiv \rho_\gamma/\rho_c = 2.47\times10^{-5}h^{-2}$.
If the extra energy density component has stayed radiation-like until today,
its ratio to the critical density, $\Omega_x$, is given by
\beq
h^2   \Omega_x \equiv h^2\frac{\rho_x}{\rho_c} = 5.61\times10^{-6}\DNeff
\eeq
today\footnote[1]{
To be more precise,
the neutrinos have not completely decoupled by the onset of
electron-positron annihilation, so that some entropy does leak to the neutrino
component.  This effect, plus finite-temperature QED corrections,
can be represented by using $\Neff = 3.04$ after $e^\pm$-annihilation
in the $\DNeff = 0$ case\cite{N304,Mangano}.
Mangano \etal\cite{Mangano} have considered the effect of extra relativistic
species on this.  Their numerical results (for $\DNeff = 0$--$1$) can be
approximated by $\Neff = 3.0395 + (1-0.0014)1.0074\DNeff$ after
$e^\pm$-annihilation,
where the part $3.0395-0.0014\times1.0074\DNeff$ is in neutrinos
and $1.0074(1-0.00025\DNeff)\DNeff$ is in the extra species\cite{Pastor}.
This assumes that the extra energy density
represented by $\DNeff$ has completely decoupled before $e^\pm$
annihilation.  Here we have defined $\DNeff$ as $\Neff-3$
before $e^\pm$-annihilation,
i.e., $\DNeff$ is a constant parameter, whereas $\Neff$ changes in the
$e^\pm$ annihilation.
Then we have $\rho_x = 0.2288(1-0.00025\DNeff)\DNeff\rho_\gamma$ after
the $e^\pm$ annihilation and
$\Omega_x = (1-0.00025\DNeff)5.653\times10^{-6}h^{-2}\DNeff$ today.
Since antimatter annihilation effects in our ABBN scenario take place
in part during the same period as $e^\pm$-annihilation, the quantity
$\DNeff$ constrained by our calculations is intermediate between
$\DNeff$
and $\Neff-3$ after $e^\pm$-annihilation.
We ignore these small effects in the present work.},
where $h$ is the Hubble constant in units of $100$~km/s/Mpc.

\subsection{Standard BBN and extra relativistic species}

The BBN isotope most sensitive\footnote[2]{The
magnitude of the effect on an isotope is considered in terms of
how accurately
its primordial abundance can be determined from observations.}
to the expansion rate is $\UHe$.  The
$\UHe$ yield depends on the number of available neutrons, or on the
number ratio
$n/p$ of neutrons and protons.  This, in turn,
is determined by the competition between the expansion rate and the rates
of the weak reactions which convert neutrons into protons before the
onset of the strong nuclear reactions which produce the isotopes.  Extra energy
density speeds up the expansion rate so that more neutrons remain available
for $\UHe$ production.

In the usual terminology, standard BBN (SBBN)
has $\DNeff = 0$, or $\Neff = N_\nu = 3$.
Then SBBN is a one-parameter theory, with the present baryon-to-photon ratio
\beq
    \eta \equiv 10^{10}\et \equiv \frac{n_b}{n_\gamma}
\eeq
the only free parameter.  It is related to $\Omega_b$, the fraction of critical
density in baryonic matter today, by
\beq
    \Omega_bh^2 = 3.65\times10^{-3}\et.
\eeq
For the purpose of the present discussion we expand the concept of SBBN by
allowing the presence of extra radiation-like energy density, so that
SBBN has two parameters, $\eta$ and $\Neff$.

Since $Y_p$, the primordial $\UHe$ mass fraction, is an increasing function of
both $\Neff$ and $\eta$ in SBBN,
the effect of a larger $\Neff$ can be compensated by a {\em lower}
$\eta$ to keep $Y_p$ in the observed range.

The effect of the expansion rate on the other isotopes, $\D$, $\EHe$,
and $\ZLi$ can be divided into two parts:
\begin{enumerate}
\item The higher $n/p$ ratio leads to an increase in the yield of the
other isotopes too.
\item The yields of these other isotopes are also sensitive
to the time scale while these isotopes are produced by the strong
nuclear reactions.  The effect of less time can be compensated by increasing
the reaction rates by increasing the baryon density.
\end{enumerate}
These effects work in the same direction for $\D$ and for $\EHe$, but (for
$\et\gtrsim3$) in the opposite direction for $\ZLi$, for which the second
effect is stronger.  Thus the net effect on the other isotopes is to shift the
agreement with observations towards {\em higher} $\eta$.
This soon leads to conflict
with $Y_p$, constraining $\DNeff$.

Copi, Schramm, and Olive\cite{Copi} concluded that $\Neff<4$ is a conservative
upper limit.  With various observational constraints on the primordial
abundances, Lisi, Sarkar, and Villante\cite{Lisi} obtained 95\% confidence
  limits in the
range $\Neff = 2$--$4$.  Using (probably prematurely) tight observational
constraints, $\DH = 3.4\pm0.25\times10^{-5}$ and $Y_p = 0.244\pm0.002$,
and the prior $\Neff \geq 3.0$, Burles \etal\cite{BurlesNeff} get
a tight upper limit $\Neff < 3.20$ (95\% confidence).  Cyburt,
Fields, and Olive\cite{Cyburt} consider various fixed values of $\eta$ (looking
forward to the future precision determinations of $\eta$ from the CMB
anisotropy) and obtain the 95\% confidence limits
$\Neff \leq 3.6 (3.0)$ for $\et = 5.8$
and $\Neff \leq 4.1 (3.9)$ for $\et = 2.4$
with (without) the prior $\Neff \geq 3.0$.

To summarize, the obtained SBBN upper limits are
\beq
    \DNeff \leq 0.2\ldots1.0
\label{sdn}
\eeq
depending on the observational constraints chosen.  Translated into an
upper limit to $\Omega_x$ they become
\beq
    h^2\Omega_x \leq 1.1\ldots5.6\times10^{-6}.
\label{somn}
\eeq

\section{BBN with matter--antimatter domains}

BBN in the presence of antimatter regions has been
discussed in \cite{ABBN,KS}.
In this scenario the baryon-to-photon ratio
is not only inhomogeneous but also not positive definite.
In spite of this the Universe is not matter--antimatter
symmetric and the
antimatter regions are small enough to be completely annihilated well
before recombination, so that they escape the CMB spectral
distortion bound\cite{CMBdis}.
This type of inhomogeneities can come from different
baryogenesis scenarios \cite{antimatter}.

The time when most of the annihilation takes place is determined by the size of
the antimatter regions, and this, in its
turn determines the nature of ABBN.
Two different cases can be distinguished,
depending on whether the annihilation takes
place before or after $\UHe$ is formed in nucleosynthesis.  The relevant
physics is completely different in the two cases.

The case of interest here is the one with the smaller distance scale,
where the antimatter region is annihilated before significant amounts of $\UHe$
is made.  This corresponds to antimatter regions of typical radius
$r_A$ between
$10^5$~m and than $10^7$~m. (We give all distances as comoving at $T=1$~keV.)
If the antimatter regions
are even smaller they annihilate before neutrino decoupling and have no effect
on BBN.  The matter and antimatter are mixed by (anti)neutron diffusion.  More
neutrons than protons are annihilated and thus the $n/p$ ratio is reduced
compared to SBBN.  Thus we get less $\UHe$.

This reduction in $Y_p$ due to neutron annihilation can be used to
cancel the increase in $Y_p$ due to a large $\DNeff$.
Since both work by affecting the $n/p$
ratio, the $n/p$ part of the effect on the other isotopes is cancelled also,
and one is left with the effect of the speed-up on the strong reactions.
This shifts the agreement with observed abundances
towards larger $\eta$ for large $\DNeff$.

For calculations (see\cite{KS} for details),
we assume an idealized geometry, where the antimatter regions
are spherical with comoving radius $r_A$ and have homogeneous
antimatter density
equal to the matter density outside the region.  With this
idealization, the ABBN scenario introduces two new parameters, $r_A$ and $R$,
where $R$ is the antimatter/matter ratio in the universe.  We assume $R<1$
so that matter is left over after the antimatter regions have annihilated.
Together $r_A$ and
$R$ determine the number density of the antimatter regions (or the typical
distance between neighboring antimatter regions).

\begin{figure}[tbh]
\epsfxsize=8cm
\epsffile{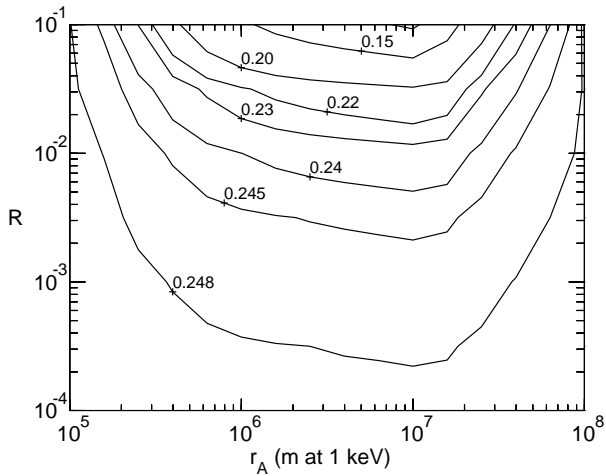}
\caption[a]{\protect
The yield of $\UHe$ as a function of the
antimatter/matter ratio $R$ and the radius $r_A$ of the antimatter
regions, for $\et=6.0$. The SBBN result, which is
approached in the lower left corner, is $Y_p = 0.2484$.
}
\label{fig:he4yield}
\end{figure}

In Fig.~\ref{fig:he4yield} we show the dependence of $Y_p$ on $r_A$ and $R$ for
$\et = 6$ and $\Neff = 3$.  We get the maximum effect for
$r\sim10^6$--$10^7$~m, when most of the annihilation takes place just before
$\UHe$ is made.  For fixed $r_A$ the effect is roughly proportional to $R$.
For $r\lesssim 10^7$~m, the effect on the other isotopes is also via the $n/p$
ratio and thus relatively small.

For $r>10^7$~m, some of the annihilation
takes place after $\UHe$ formation, leading to significant production of $\EHe$
and $\D$.  Because of $\EHe$ overproduction, this region is not interesting
for allowing larger $\Neff$.

%

\section{ABBN and additional relativistic species}

The observational constraints on the primordial abundances
of the light elements are a matter of some controversy\cite{BBNobs}.
We follow Kneller \etal\cite{Kneller}
in adopting the relatively generous constraints
\bea
   0.23 \leq & Y_p & \leq 0.25 \\
   2\times10^{-5} \leq & \DH & \leq 5\times10^{-5}  \\
   1\times10^{-10} \leq & \ZLiH & \leq 4\times10^{-10}.
\eea
We search for the region in the parameter space which satisfies these
constraints.  In SBBN these constraints lead to the upper limit
$\Neff \leq 3.4$.
For $\Neff =3$ the same constraints limit $\eta$ to the range
$4.2 \leq \et \leq 5.9$.

For $r_A<10^7$~m, the effect of the antimatter regions on nucleosynthesis
is mainly due to the reduction of the $n/p$ ratio. As shown below,
the dependence of the light element yields
on $r_A$ and $R$ can be combined into a dependence on a single
parameter, the $n/p$ ratio at the onset of nucleosynthesis.
This redundancy of parameters allows us to cover the whole
phenomenology by varying three of the four
parameters only.  For most of our computations we kept
the antimatter radius fixed at
$r_A=10^{6.9}$~m, in order to get the maximal reduction in $n/p$
for a given value of $R$. This
left us with three parameters
$\eta$, $\Neff$, and $R$. We calculated the ABBN yields of light
elements in this 3-dimensional parameter space and compared them to the
observational constraints.

\begin{figure}[tbhp]
\epsfxsize=8cm\epsffile{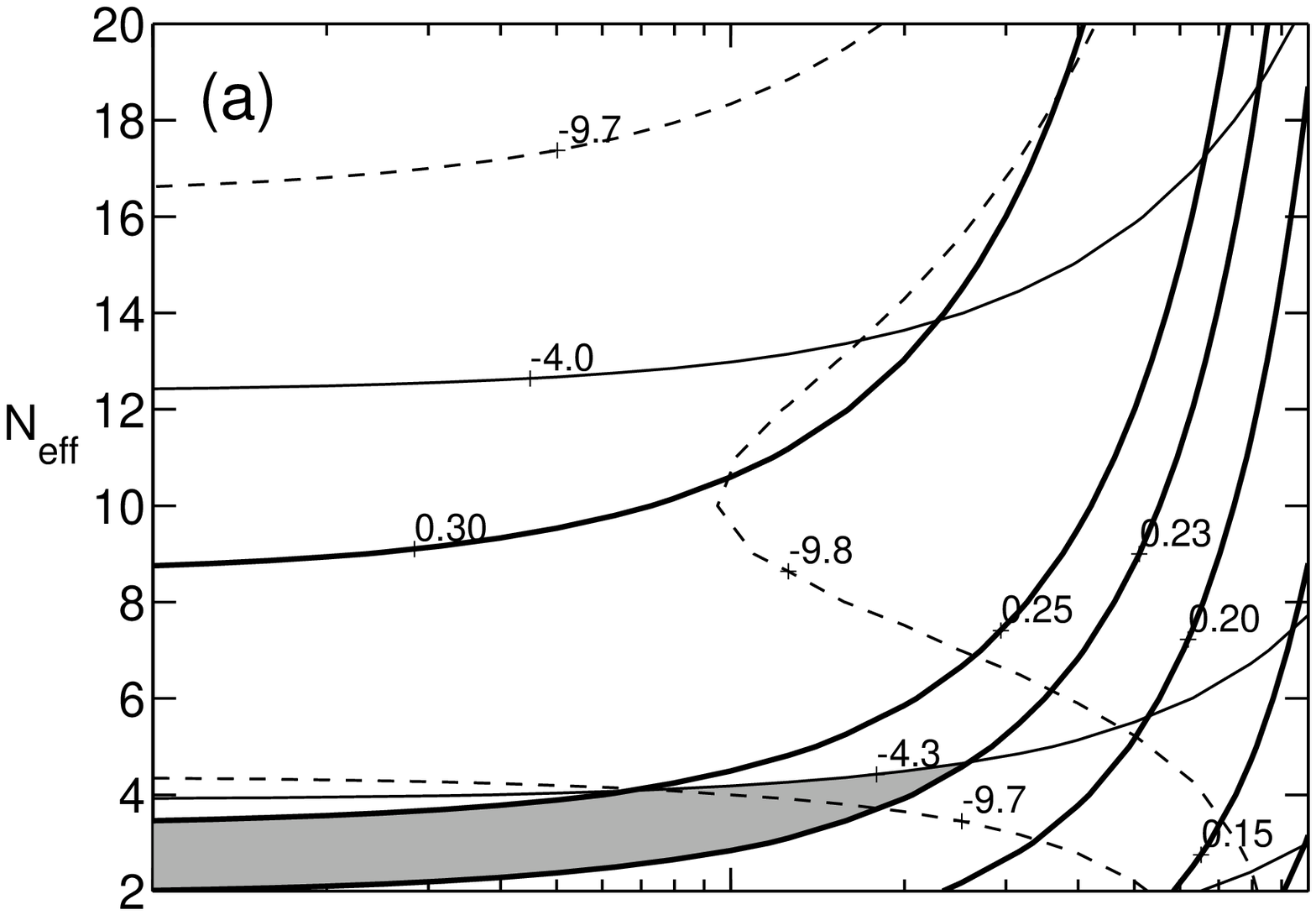}
\epsfxsize=8cm\epsffile{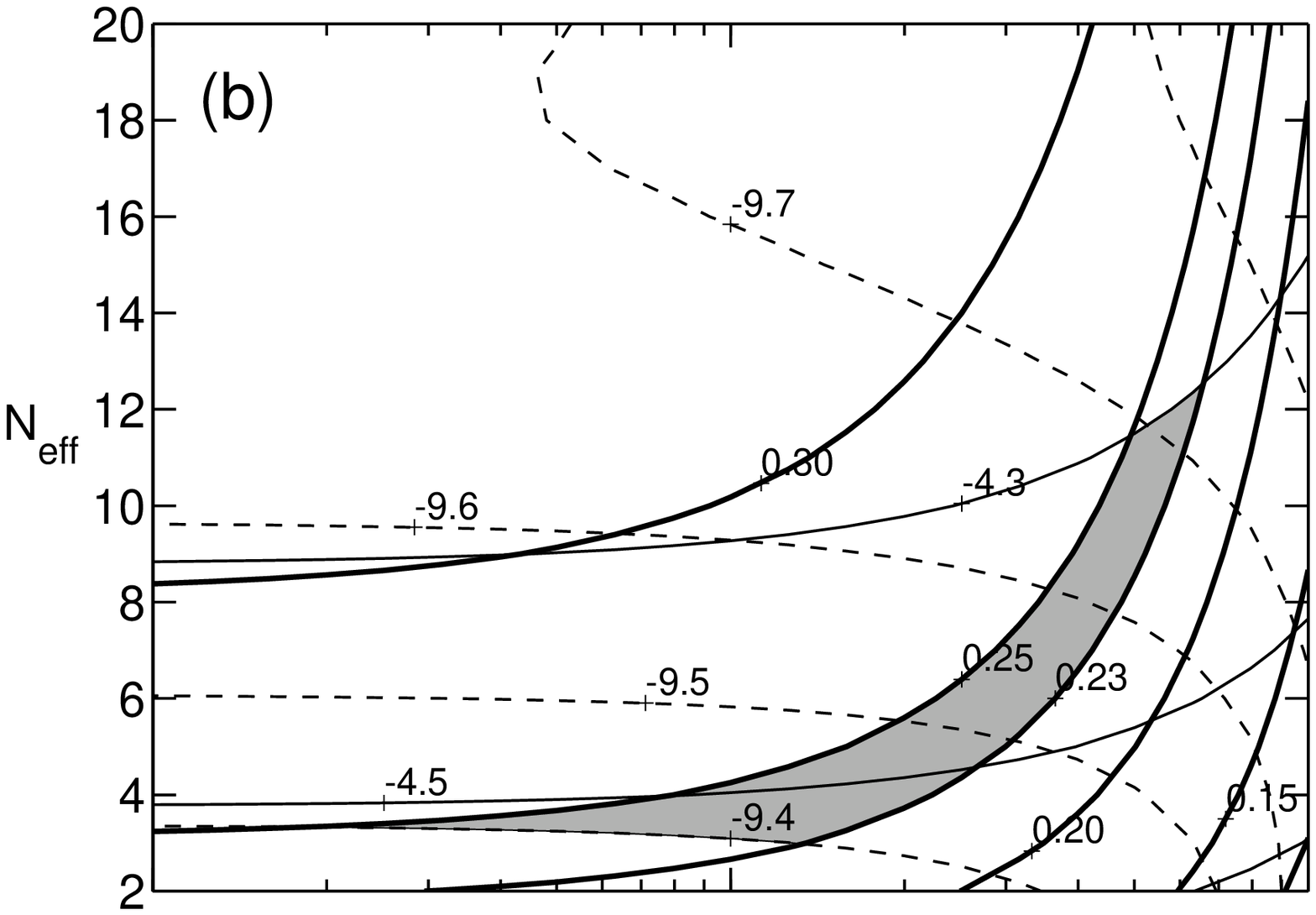}
\epsfxsize=8cm\epsffile{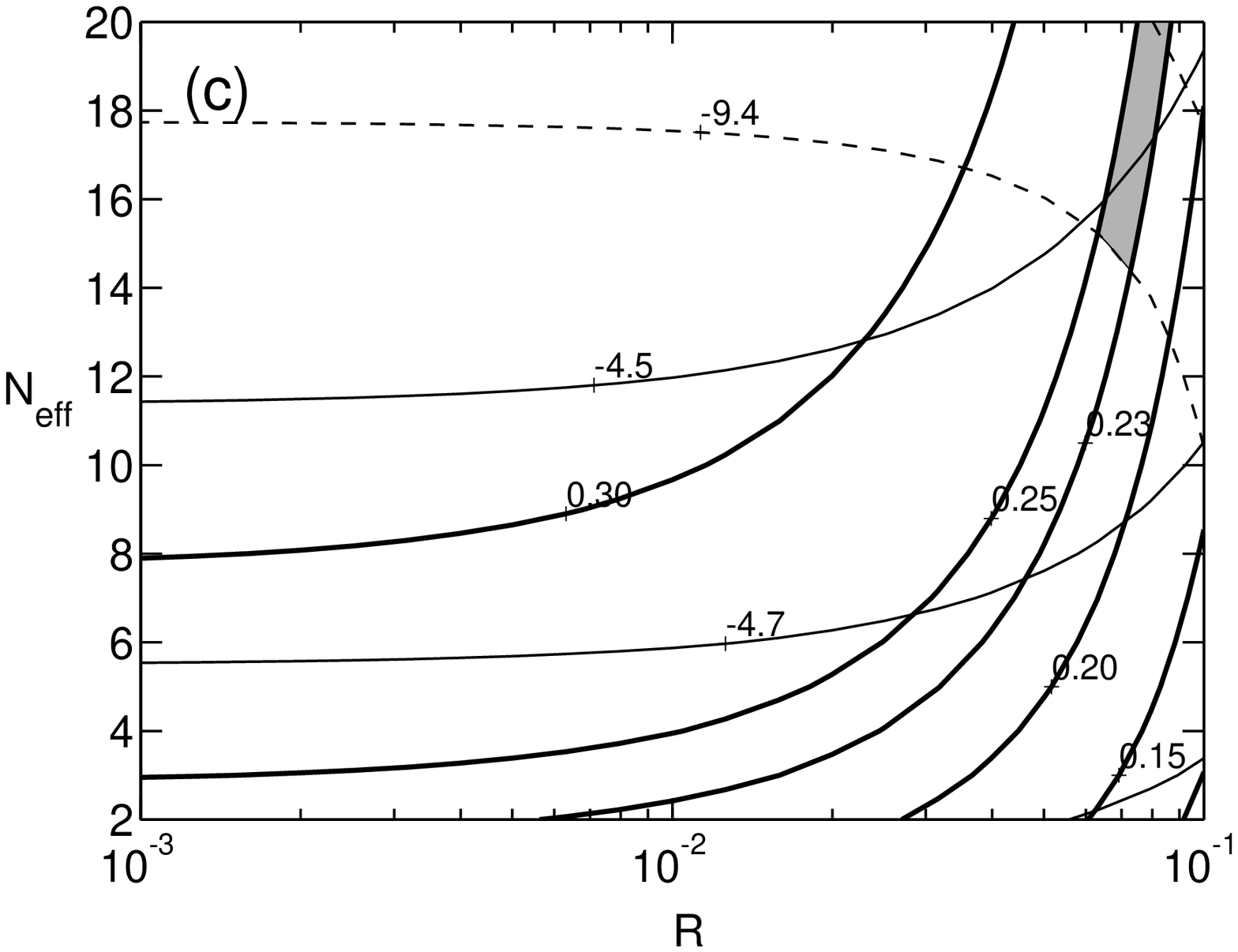}
\caption[a]{\protect
The light element yields in ABBN for
a fixed $r_A = 10^{6.9}$~m and $\et = 4.5$
(a), $6.0$ (b), and $9.0$ (c), as a function of the two remaining
parameters, $R$ and $\Neff$.  We show contours for $Y_p$ ({\em thick
solid lines}),
$\log_{10}\DH$ ({\em thin solid lines}),
and $\log_{10}\ZLiH$ ({\em dashed lines}).  The shaded region is
allowed by our adopted
observational constraints.
}
\label{fig:fixeta}
\end{figure}

In Fig.~\ref{fig:fixeta} we show slices of this 3-dimensional parameter space
for fixed values of $\eta$. For a given $\eta$, the upper limit on
$\Neff$ comes from a
combination of the upper limit on $\DH$ and the lower limit on $\UHe$.

\begin{figure}[tbhp]
\epsfxsize=8cm\epsffile{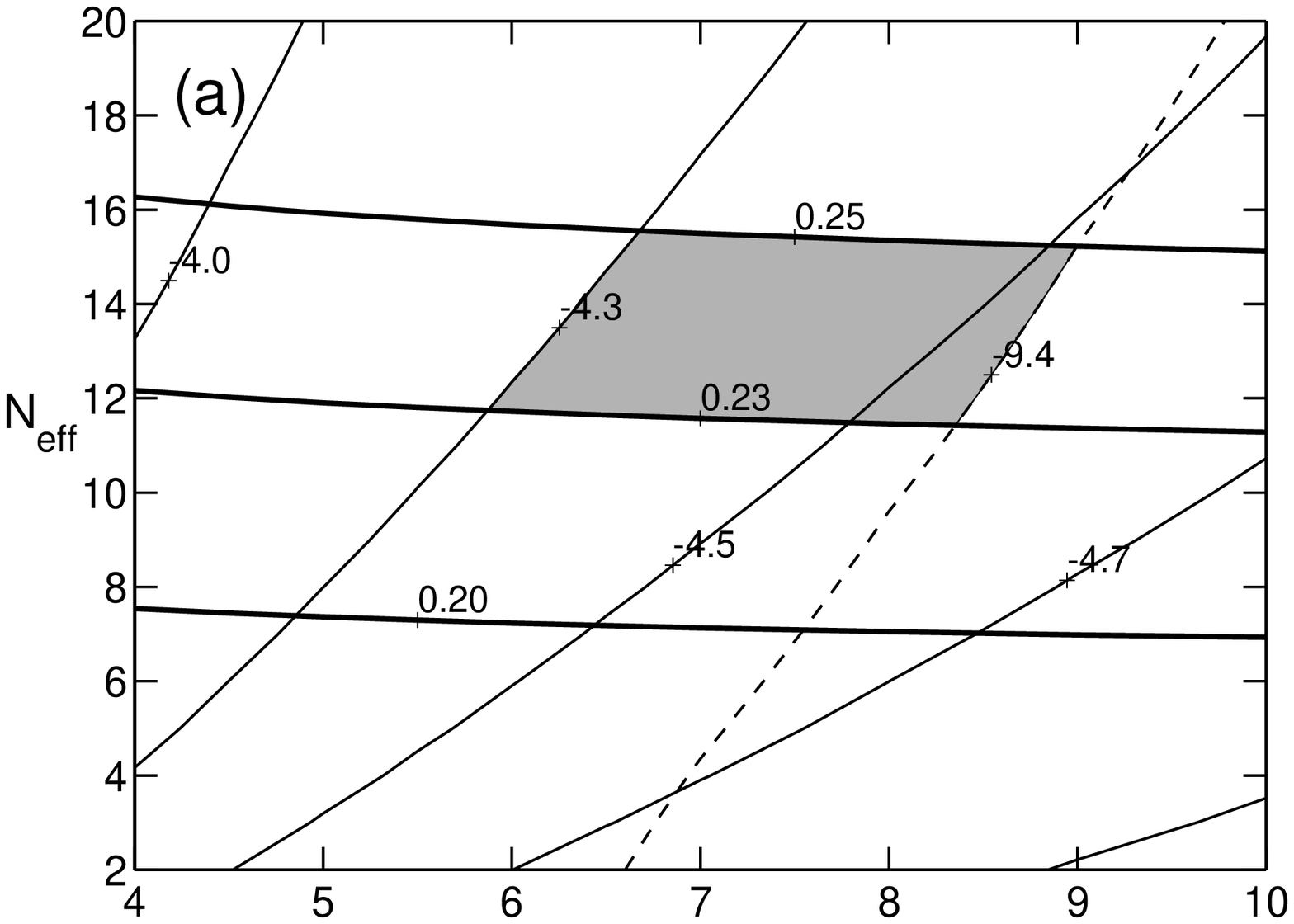}
\epsfxsize=8cm\epsffile{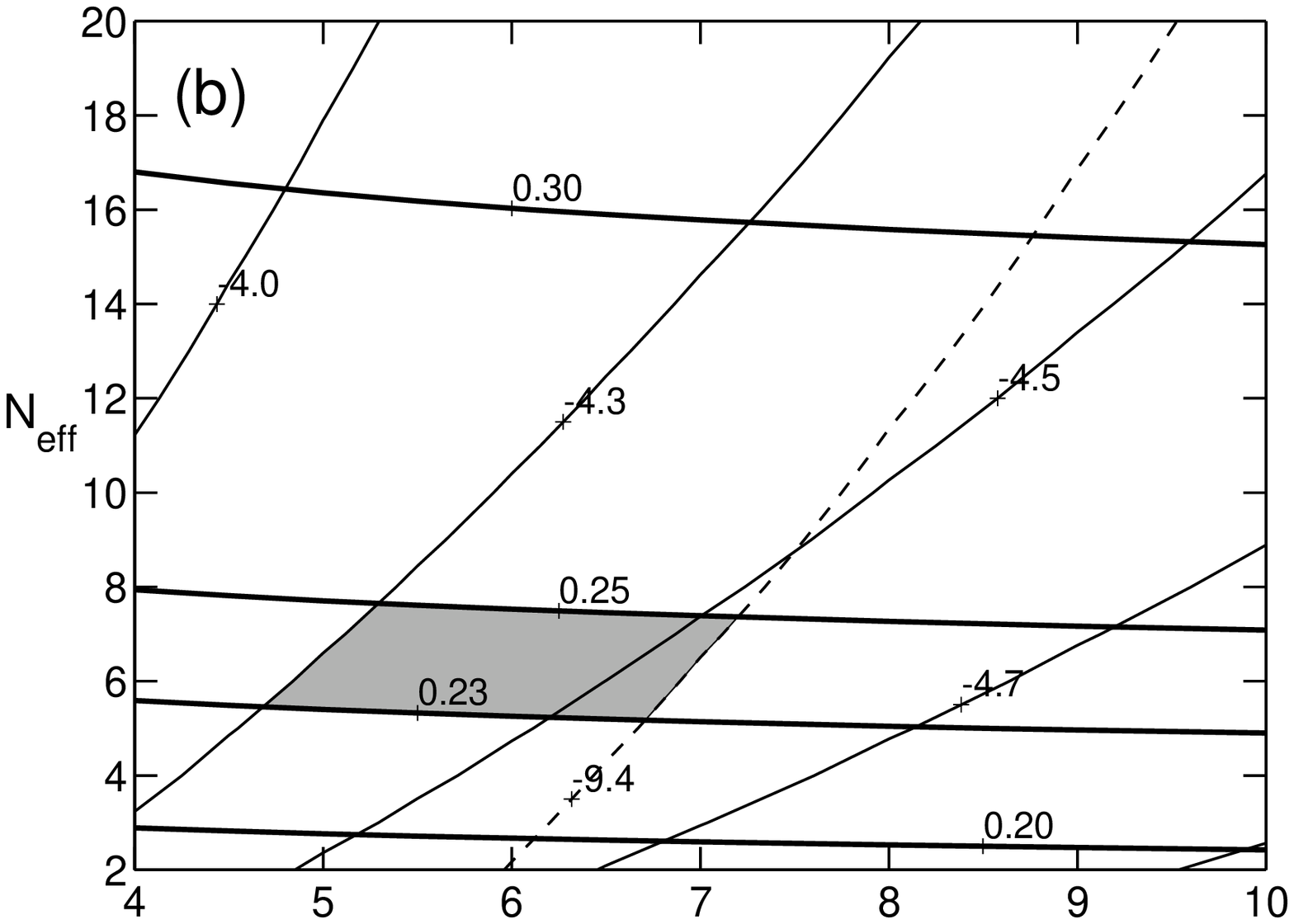}
\epsfxsize=8cm\epsffile{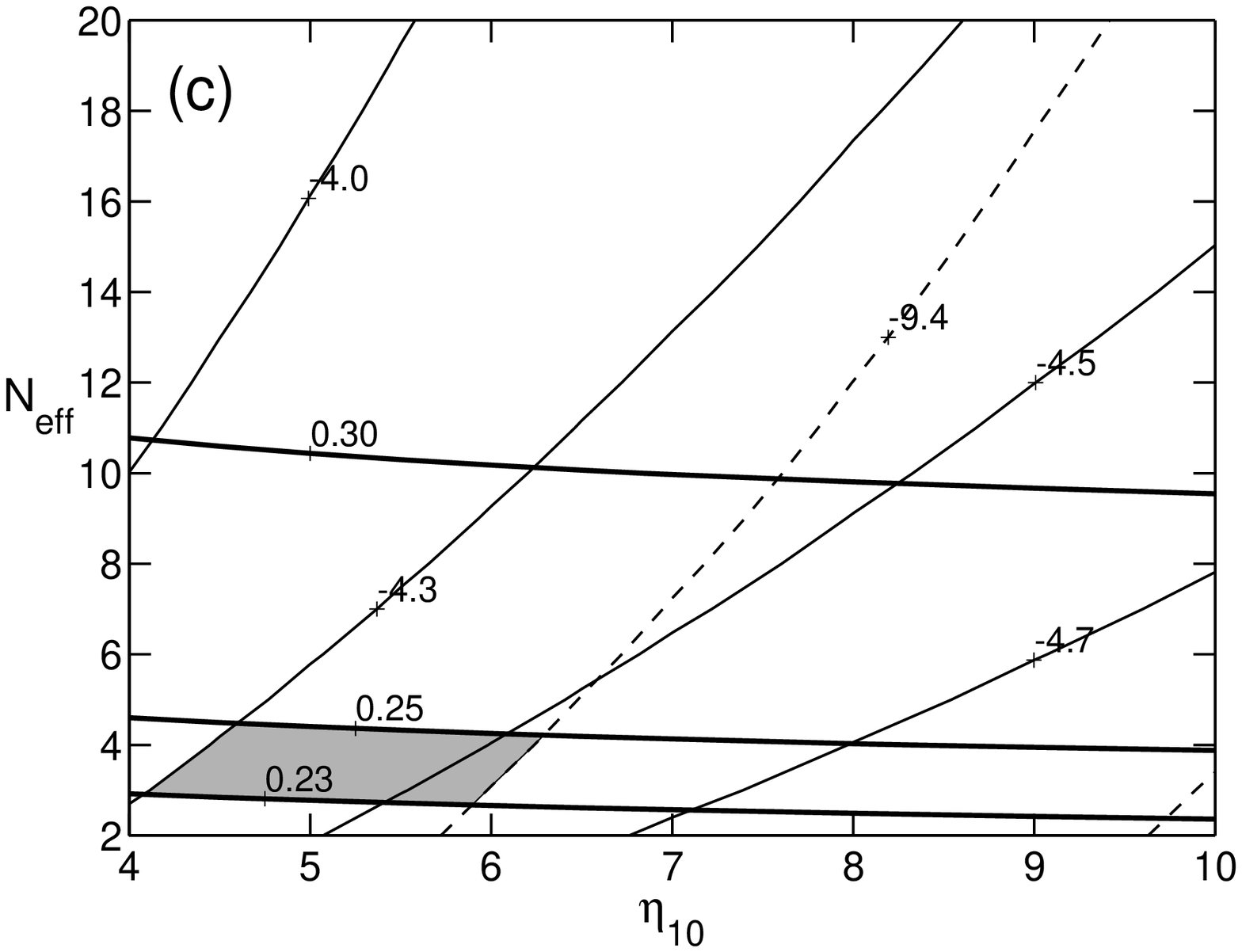}
\caption[a]{\protect
The light element yields in ABBN for $r_A = 10^{6.9}$~m and
antimatter/matter ratio $R = 10^{-1.2}$ (a), $10^{-1.5}$ (b), and $10^{-2}$
(c), as a function of $\eta$ and $\Neff$. The line types and shading
are the same as in
Fig.~\ref{fig:fixeta}. }
\label{fig:fixR}
\end{figure}

In Fig.~\ref{fig:fixR} we show
slices along another dimension, for fixed values of $R$. The allowed
$\Neff$ increases with
increasing $R$. Simultaneously the allowed range in $\eta$ shifts to
higher values.

%

%

\begin{figure}[tbhp]
\epsfxsize=8cm\epsffile{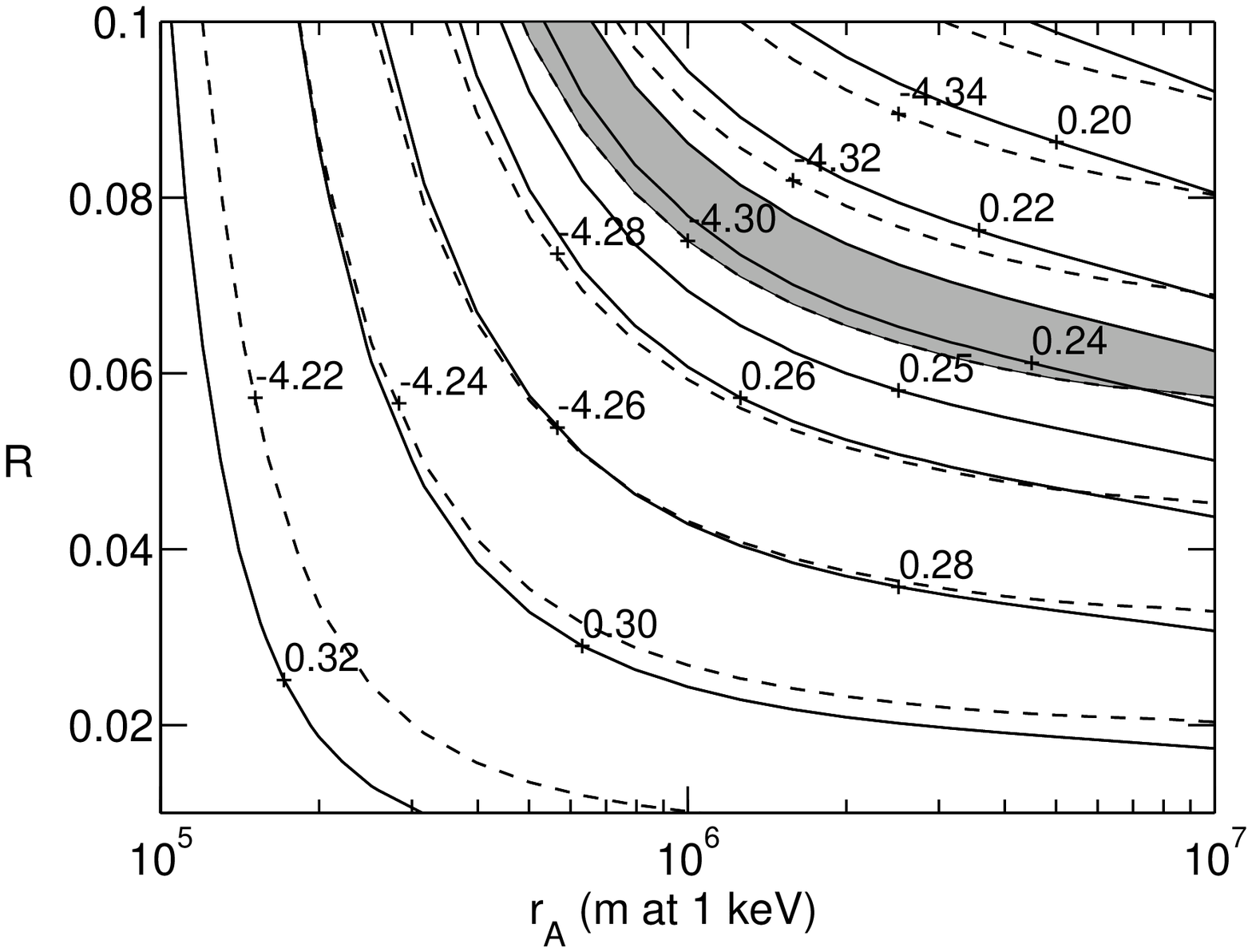}
\caption[a]{\protect
The effect of the antimatter radius $r_A$ on the production of $\UHe$ and
D.  We show contours of
$Y_p$ ({\em solid lines}) and
$\log_{10}\DH$ ({\em dashed lines}) on the
$(r_A,R)$-plane, for $\et = 6.0$ and $\Neff = 12$.  The contours are almost
parallel, showing that the yields depend on $R$ and $r_A$ only
through a single combination.
}
\label{fig:r}
\end{figure}

Fig. \ref{fig:r} illustrates the redundancy of parameters $R$ and
$r_A$. We show the yields of
$\UHe$ and deuterium as a function of $r_A$ and $R$, for fixed
$\et=6.0$ and $\Neff=12$. The value of $\Neff$ was chosen close to
its upper limit for this value of $\et$
(see Fig.~\ref{fig:fixeta}b).
Reducing $r_A$ moves the allowed region to higher values of $R$,
but the width of the region remains nearly the same.  Thus the effect of a
smaller $r_A$ can be compensated by a larger $R$.  The contours of
$Y_p$ and $\DH$ run almost parallel, so that we get essentially
the same results for different combinations of $r_A$ and $R$, which
lie along these contours.  This shows that the
light element yields
depend on a single function of $R$ and $r_A$.

%

%
\begin{figure}[tbhp]
\epsfxsize=8cm\epsffile{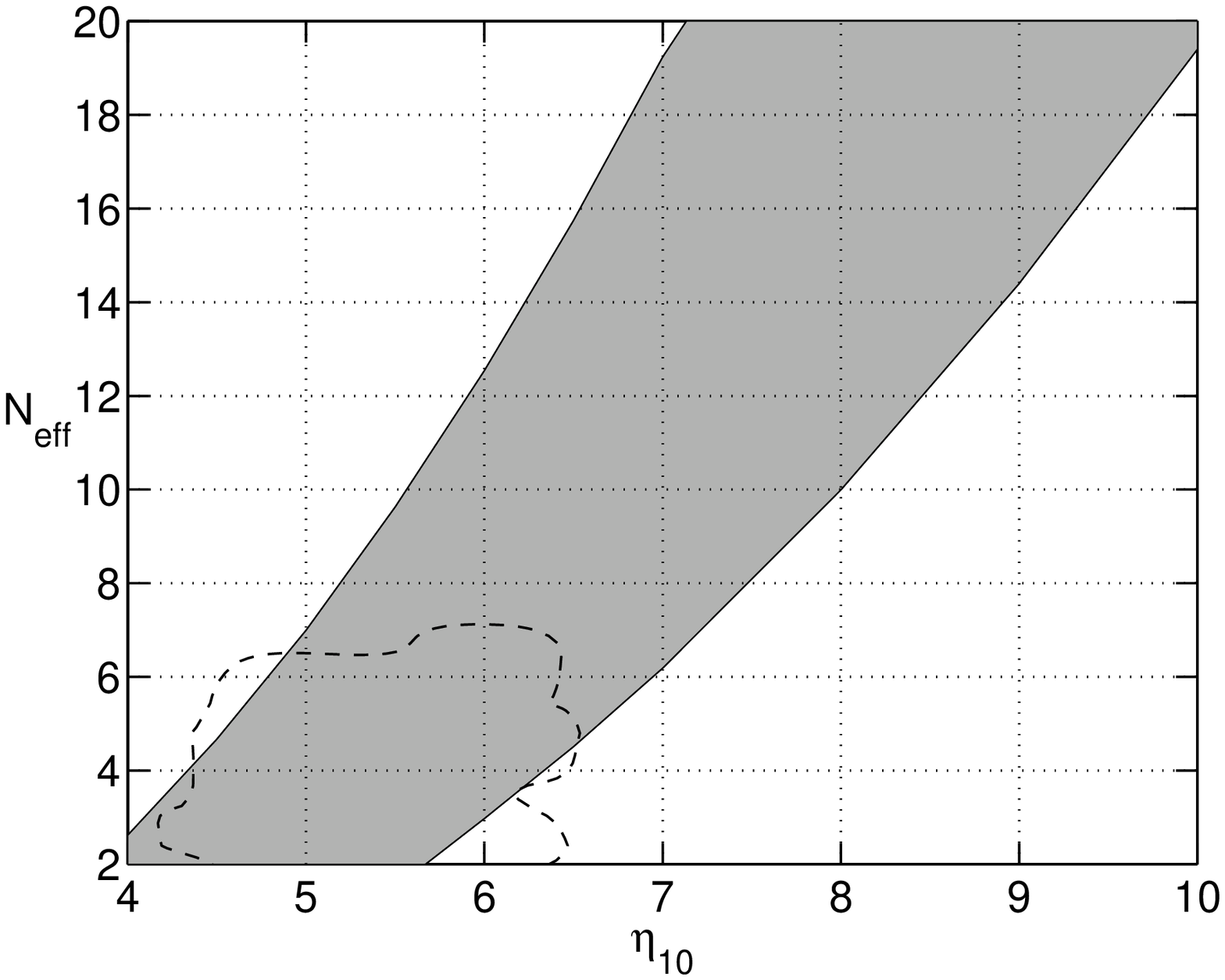}
\caption[a]{\protect
The combined allowed region in $(\eta,\Neff)$. The shaded region
indicates values of
$\eta,\Neff$ which are allowed for {\it some} value of the
antimatter/matter ratio $R$.
The dashed line is the CMB+SNIa constraint from\cite{Hansen}.}
\label{fig:allowed}
\end{figure}

In Fig.~\ref{fig:allowed} we show the projection of the allowed
region in the $(\eta,\Neff,R)$
space onto the $(\eta,\Neff)$-plane.
We see that very large amounts of extra radiation energy can be accommodated
in ABBN.  A large $\Neff$ requires a somewhat larger $\eta$.
For, e.g, $\et=6.0$ we obtain $3\lesssim\Neff\lesssim12.5$.
Even values $\Neff > 20$ cannot be ruled out on the basis of nucleosynthesis.
These large values require $\et > 7$.

If we accept the possibility of ABBN, $\Neff$ is no longer constrained by BBN,
but is limited by other observational constraints.

\section{Comparing ABBN to DBBN}

\subsection{Degenerate BBN}

It is interesting to compare these results to another NSBBN scenario,
degenerate BBN (DBBN) \cite{otherDBBN,Kang},
which is also very effective in reducing the $\UHe$ yield.
In SBBN the neutrino chemical potentials are assumed to be small, so that there
are almost equal numbers of neutrinos and antineutrinos.  In DBBN
this assumption is lifted.  The degeneracy parameters
\beq
    \xi_i \equiv \frac{\mu_{\nu_i}}{T_\nu}, \qquad (i = e,\mu,\tau),
\eeq
stay constant as the temperature falls, as long as the neutrinos are
relativistic.
A nonzero $\xi_i$ increases the energy density in
neutrinos of type $i$ by
\beq
    \DNeff = \frac{30}{7\pi^2}\xi_i^2 + \frac{15}{7\pi^4}\xi_i^4.
\eeq
Degenerate neutrinos are thus one possibility for a nonzero (positive)
$\DNeff$.

If the degeneracy is in electron neutrinos, a much stronger effect is that it
shifts the balance of the weak reactions between neutrons and protons.  The
equilibrium ratio is
\beq
    \left(\frac{n}{p}\right)_{\rm eq} = e^{-Q/T-\xi_e},
\eeq
where $Q \equiv m_n - m_p$.  Thus a positive $\xi_e$ reduces $n/p$ and thus
$Y_p$, while a negative $\xi_e$ increases $n/p$ and $Y_p$.  This effect is much
larger than the speed-up effect from the $\DNeff$ due to $\xi_e$.  One can have
both $\DNeff$ and $\xi_e$ as free parameters in this scenario by allowing
the other $\xi_i$ to have much larger (absolute) values than $\xi_e$
(this possibility might, however,
be excluded by neutrino oscillations\cite{Dolgov02}),
or by assuming some other extra radiation-like energy density contribution.

In DBBN one can compensate the increase in $Y_p$ due to a large $\DNeff$ by
a reduction due to a positive $\xi_e$.  Like in ABBN, one then compensates
for the speed-up effect on the other isotopes by increasing $\eta$.
However this effect on these other isotopes is relatively small, and one
can go to
quite large $\DNeff$ with a relatively small shift in $\eta$.

Considering BBN alone, the limit to $\DNeff$ is practically removed.  Combining
DBBN limits with other cosmological bounds on the density and expansion rate of
the universe, Kang and Steigman\cite{Kang} obtained (in 1992) the limits
  $-0.06 \leq
\xi_e \leq 1.1$, $|\xi_{\mu,\tau}| \leq 6.9$, which correspond to $\DNeff \leq
71$, and $2.8\leq\et\leq19$. The other cosmological constraints have since
gotten tighter (see Sec.~\ref{sec:diffcon} below).

\subsection{ ABBN versus DBBN}

The DBBN allowed region in $\eta$ and $\DNeff$
(see, e.g., Fig.~2 of\cite{Esposito}, Fig.~5 of\cite{Kneller}, or Fig.~1
of\cite{Hansen}) is remarkably similar
to that in ABBN (Fig.~\ref{fig:allowed}).  We can understand this as follows.

In both cases the nonstandard feature affects nucleosynthesis mainly
by reducing the $n/p$ ratio, with practically
arbitrarily large reductions possible. Thus the nonstandard parameters,
$(r,R)$ in the case of ABBN and $\xi_e$ in the case of
DBBN, can be mapped onto a reduction factor in
$n/p$.  The rest of the game is then the same:  $\DNeff$ can now be increased
to bring the $n/p$ ratio back up to give the observed value of $Y_p$; this has
then a small effect on the other isotopes by shortening the time available for
nuclear reactions, which has to be compensated by an increase
in $\eta$.

\section{Different constraints on additional
relativistic species}
\la{sec:diffcon}

A large radiation energy density delays (i.e., moves to lower temperature) the
transition from radiation domination to matter domination in the expansion law
of the universe, and shortens the time scale of the radiation-dominated epoch.
Since the radiation-dominated expansion law allows for logarithmic growth of
density perturbations only, the large-scale structure of the universe
constrains the radiation component.  The effect of the radiation component in
the expansion law shows also as an integrated Sachs-Wolfe (ISW) effect in the
CMB angular power spectrum.  Hannestad\cite{Hannestad}
used this to obtain an upper limit $\Neff < 17$ (95\% confidence).

In connection with a NSBBN scenario one can go further and combine the NSBBN
constraints on $(\eta,\Neff)$ with the CMB constraints on $(\eta,\Neff)$.
This has been done for DBBN in \cite{Orito,Esposito,Kneller,Hansen,Lesg}.
These results will apply also to ABBN,
since the allowed $(\eta,\Neff)$ region is essentially the same in both NSBBN
scenarios.   This NSBBN allowed region leans towards higher $\eta$ for high
$\Neff$,
whereas the CMB allowed region leans (slightly) towards lower $\eta$
for high $\Neff$.  The regions overlap for $\DNeff\sim$~few, but for the
highest
$\DNeff$ allowed by CMB, the CMB seems to require a lower $\eta$ than these
NSBBN
scenarios allow.  Thus the combined constraint on $\DNeff$ may be lower than
that from CMB alone.  Hansen \etal\cite{Hansen} use a different set of CMB
data, favoring somewhat lower $\eta$,
than Kneller \etal\cite{Kneller} use, and get therefore tighter combined upper
limits for $\DNeff$.

Since the CMB spectrum depends on a large number of other cosmological
parameters, these results are sensitive to the assumed priors on these other
parameters.  Kneller \etal\cite{Kneller} find that values of $\DNeff$ larger
than 10 are compatible with combined CMB and DBBN constraints, but when they
include
a prior on the matter density parameter, $\Omega_M \leq 0.4$, and on the age of
the universe, $t \geq 11$~Gyr, they get a tighter upper limit, $\DNeff
\lesssim6$.
Hansen \etal\cite{Hansen} get  $\DNeff\leq5$ with a prior on the age,
  $t > 11$~Gyr,
and $\DNeff\leq4$ (both at 95\% CL) with a prior on the parameters
$\Omega_M$ and $\Omega_\Lambda$ based on the supernova Ia data
(see Fig.~\ref{fig:allowed}).

As emphasized  by Hansen \etal\cite{Hansen}, the determination of $\eta$ from
the CMB data relies on the heights of both the second and the third acoustic
peaks in the CMB angular power spectrum, and, especially for the third peak,
further data is needed to check on possible systematic effects.  Therefore
these combined CMB+NSBBN upper limits to $\DNeff$ should be taken as
preliminary.

At present, we can conclude that values of $\DNeff$ as large
as $4$ are acceptable, and values somewhat larger, $\DNeff = {\cal{O}}(10)$ may
not yet have been ruled out.  Thus, allowing for nonstandard BBN, like ABBN,
raises the upper limit to the present fraction of extra radiation-like
energy density from
\beq
    h^2\Omega_x \leq 1.1\ldots5.6\times10^{-6}
\label{aom1}
\eeq
to
\beq
    h^2\Omega_x \leq 2.3\ldots8\times10^{-5},
\label{aom2}
\eeq
an increase by an order of magnitude.

\section{ABBN and stochastic backgrounds of relic
gravitational waves}

The supplementary radiation allowed either by ABBN or by DBBN
may be the consequence of a primordial background of gravitational radiation.
The present critical fraction of energy density stored in relic gravitons
\beq
   h^2   \Omega_{\rm GW}(t_0) \equiv h^2\frac{\rho_{\rm GW}}{\rho_c}
   = 5.6\times10^{-6}\Delta  N_{\rm eff}.
\label{g1}
\eeq
depends upon $\Delta N_{\rm eff}$ whose range of variation
can  be translated
into constraints on the energy density of
relic gravitational waves produced prior to BBN.
The critical fraction
of spectral energy density per logarithmic interval of frequency
\beq
\Omega_{\rm GW}(\nu,t_0) = \frac{1}{\rho_{\rm c}}
\frac{d \rho_{\rm GW}}{ d\ln{\nu}}.
\label{sp}
\eeq
is often introduced for practical purposes. Eqs. (\ref{sdn})--(\ref{somn})
(obtained in the context of SBBN) and Eqs. (\ref{aom1})--(\ref{aom2})
(obtained in the context of ABBN) can then be interpreted, in light
of Eq. (\ref{g1}), as an upper
limit on the energy density presently stored in relic gravitons.
Heeding experimental observations, the  correlation of resonant mass
detectors or of wide band interferometers can also provide upper
bounds on the critical fraction of energy density stored in relic gravitons.
The sensitivity
of operating resonant mass detectors is not yet able to
probe possible signals compatible with  Eqs.~(\ref{aom1})--(\ref{aom2})
and (\ref{g1}).
Forthcoming generations
of resonant mass and interferometric detectors are expected to improve
their sensitivity in $h^2\Omega_{\rm GW}$. The bounds
coming from ABBN and from direct searches of stochastic gravitational wave
backgrounds will represent two complementary sets of constraints on the same
quantity.

\subsection{GW detectors and BBN bound}
Various resonant mass
detectors  are now operating
\cite{allegro,auriga,explorer,nautilus,niobe}.
In \cite{astone}, the first
experiment of cross-correlation between two cryogenic detectors
has been reported  with the purpose
of giving an upper limit on  $h^2 \Omega_{\rm GW}$. The two detectors are
Explorer \cite{explorer} (operating in CERN, Geneva) and Nautilus \cite{nautilus}
 (operating in Frascati,
near Rome). Previous experiments giving upper limits on $h^2 \Omega_{\rm GW}$
used room temperature detectors. The Rome group obtained then an
upper limit $h^2 \Omega_{\rm GW} < 60$ at a frequency of
roughly $905$ Hz. The limit is a result of cross-correlation between the two
detectors (located at a distance of approximately $600$ km) for an integration
time of approximately 12 hours. This limit is not competitive with the BBN bound
(and also  above the critical density bound implying that $\Omega_{\rm GW}<1$).
However, by increasing the correlation time from few hours
to few months it is not unreasonable to go below one in $ h^2\Omega_{\rm GW}$.

Hollow spherical detectors
have been recently investigated \cite{fafone} as a possible tool
for the analysis
of the relic gravitational wave background. The sensitivity of two
correlated spherical detectors could be ${\cal O}(10^{-6})$ in
$h^2 \Omega_{\rm GW}$ for the frequency of resonance which lies between
200 and 400 Hz. In this case the ABBN bound and the experiment will be
certainly competitive.  Dual spherical detectors \cite{cerdonio}  may reach
a sensitivity, in $h^2 \Omega_{\rm GW}$, which is again ${\cal O}(10^{-6})$ in the kHz
region.

Consider now the case of wide band interferometers
\cite{geo,tama,virgo,ligo} whose arms  range from 400 m of TAMA
up to the 3 km of VIRGO and to the 4 km of LIGO.
The foreseen noise power spectra $S_{n}(\nu)$ are defined for $\nu$ between
a few Hz and $10$ kHz. While at small frequency the seismic noise dominates,
at higher frequencies the main source of noise is provided by the shot noise.
The signal-to-noise ratio (SNR) will be obtained by comparing the
analytic form of the noise power spectra with the possible stochastic signal
parametrized in terms of $\Omega_{\rm GW}(\nu)$.
The single LIGO
\cite{ligo} detector will be presumably sensitive
to $h^2 \Omega_{\rm GW} \sim 10^{-3}$. Two correlated interferometers
lead to a SNR
\begin{eqnarray}
&&{\rm SNR}^2 \,=\,\frac{3 H_0^2}{2 \sqrt{2}\,\pi^2}\,F\,\sqrt{T}\,\times
\nonumber\\
&&\left\{\,\int_0^{\infty}\,{\rm d} \nu\,\frac{\gamma^2 (\nu)\,
\Omega_{{\rm GW}}(\nu)}
{\nu^6\,S_n^{\,(1)} (\nu)\,S_n^{\,(2)} (\nu)}\,\right\}^{1/2}\; ,
\label{g2}
\end{eqnarray}
($H_0$ is the present value of the Hubble parameter and $F$ depends upon
the geometry of the two detectors; in the case of the correlation between
two interferometers $F=2/5$).
In Eq. (\ref{g2}), $S_n^{\,(k)} (\nu)$ is the (one-sided) noise power
spectrum of the $k$-th
$(k = 1,2)$ detector, while $\gamma(\nu)$ is the overlap reduction function
\cite{michelson,christensen,flanagan,allen}
which is determined by the relative locations and orientations
of the two
detectors. This function cuts off (effectively) the integrand at a frequency
$f \sim 1/2d$, where $d$ is the separation between the two detectors.
Eq. (\ref{g2})
assumes, as usually done, that the noises of the two detectors
are stationary, Gaussian and not
correlated.

Consider now the simple case of a flat spectrum, namely
$\Omega_{\rm GW}(\nu) = \Omega_0 \nu^{0}$.
 From Eq.~(\ref{g2}) using the appropriate
  expressions of the noise power spectra
  and of the overlap reduction functions pertaining to the LIGO and VIRGO
detectors \cite{giovannini}
we get that the sensitivity is
\begin{equation}
h^2\,\Omega_{\rm GW}\,\simeq\,1.8\,\times\,10^{-10}\;
\left(\,\frac{1\;{\rm yr}}{T}\,\right)^{1/2}\;{\rm SNR}^2
\label{qligo}
\end{equation}
for the correlation between two LIGO detectors \cite{allen,giovannini} and
\begin{equation}
h^2\,\Omega_{\rm GW}\,\simeq\,7.2\,\times\,10^{-8}\;
\left(\,\frac{1\;{\rm yr}}{T}\,\right)^{1/2}\;{\rm SNR}^2
\label{qvirgo}
\end{equation}
for the hypothetical correlation of two VIRGO detectors \cite{giovannini}.
In Eqs.~(\ref{qligo}) and (\ref{qvirgo}) $T$ is the integration time of,
for instance, one  year.

In the context of the standard BBN scenario we would be led
to exclude $h^2\Omega_{\rm GW} > 5.6 \times 10^{-6}$
as a signal of primordial origin. This bound can be
relaxed, according to Eqs.~(\ref{aom1})--(\ref{aom2})
 up to values which are ${\cal O}(10^{-4})$,
in the context of ABBN. If a signal
with  $h^2\Omega_{\rm GW}$ larger than the bound implied
by standard BBN will ever be detected, it could still be of primordial
nature. This observation may also be relevant for the bounds
on stochastic GW backgrounds obtained by a single detector, where,
$h^2 \Omega_{\rm GW}$ can be, at most ${\cal O}(10^{-4})$.

Microwave cavities have been originally proposed as GW detectors in the
GHz--MHz region of the spectrum \cite{mw1,mw2}. Their application
for high frequency gravitational wave backgrounds was theoretically suggested
in \cite{gr2}. Improvements in the quality
factors of the cavities (if compared
with the prototypes of \cite{mw1}) have been recently
achieved \cite{paco} and two
 experiments (in Italy \cite{paco} and in England \cite{bir})
are now in progress.
The sensitivity in $h^2 \Omega_{\rm GW}$ is still above one for
the experiments reported so far.
These devices could be important for the analysis of
 a frequency range where possible backgrounds generated by
collections of astrophysical sources are negligible.

\subsection{BBN bounds and theoretical models of stochastic GW backgrounds}
When the ordinary inflationary phase
(of de Sitter
or quasi-de Sitter type) is followed by a radiation epoch gravitational waves
are produced and the
theoretically estimated $\Omega_{\rm GW}(\nu) $
decreases as $\nu^{-2}$  for $10^{-18} ~h {\rm Hz} < \nu < 10^{-16} {\rm Hz}$.
This branch of the spectrum corresponds to modes leaving the horizon during
the inflationary phase and re-entering during the matter-dominated phase
\cite{rubakov,abott}.
Since $\Omega_{\rm GW}(\nu)$ is
 a  decreasing function of the present frequency $\nu$ the most
relevant bounds (determining the normalization of the spectrum) will
be the ones
coming from small frequencies (i.e. large length scales)
and, among them, the analysis
of the first thirty multipoles of temperature anisotropies in the
microwave sky. The COBE
normalization  implies, in the low-frequency part of the inflationary spectrum,
 that $h^2 \Omega_{\rm GW}(\nu) < 6.9 \times 10^{-11}$ for
$\nu \simeq 10^{-18} ~h {\rm Hz}$.

The modes re-entering during the
 radiation dominated epoch  correspond to frequencies
$10^{-16 } {\rm Hz} < \nu < 10^{-11} \sqrt{H_1/M_{P}}$ Hz,
(where  $H_1\leq 10^{-6} M_{P}$ is the curvature scale at the end of inflation).
The theoretical $\Omega_{\rm GW}(\nu)$
is either a flat (exact Harrison-Zeldovich spectrum \cite{inflsp})
or a slightly decreasing
(in the quasi-de Sitter case) function of $\nu$, i.e.  $\Omega_{\rm
GW}(\nu,t_{0}) \propto \nu^{\alpha} $ , with $\alpha \leq  0$.
Imposing the COBE normalization as illustrated above, it turns out
that
$\Omega_{\rm GW}(\nu) < 10^{-15}$ for $\nu > 10^{-16}$ Hz.
From Eqs. (\ref{qligo})--(\ref{qvirgo}) and  (\ref{g1})
it can be deduced that $\Omega_{\rm GW}<10^{-15}$ will always be compatible
with the BBN constraints and out the foreseen sensitivity of
wide band interferometers.

In quintessential inflationary models \cite{peebles},
the inflationary phase is not immediately
followed by a radiation-dominated phase, but a long stiff phase
(dominated by the kinetic energy of the quintessence field)
takes place prior to the dominance of radiation. As a consequence,
$\Omega_{\rm GW}(\nu) $ increases as $\nu $ (up to logarithmic corrections)
  for $ 10^{-3} {\rm Hz}<\nu
< {\rm GHz}$ \cite{mg}. This branch of the spectrum corresponds to
modes leaving
the horizon during the inflationary phase and re-entering during the
stiff phase.
Around the GHz, $h^2 \Omega_{\rm GW}(\nu)$ exhibits a spike.
While the
presence of the spike is a consequence of the fact that the inflationary epoch
is followed by a stiff phase, the height of the
spike is bounded by the BBN. In the case of standard BBN the
height of the spike turns out to be, at most, $0.8 \times 10^{-6}$\cite{mg}.
Using the ABBN constraint the height of the spike becomes $1.7\times 10^{-5}$.
In  pre-big bang models a similar discussion applies since
 the spectra of relic gravitons increase with frequency
\cite{mg2} and the most significant phenomenological
bounds will be the one provided by Eq.~(\ref{g1}).

\section{Conclusions}

Big bang nucleosynthesis provides the tightest constraint on the radiation-like
energy density in the universe.  However, some nonstandard BBN scenarios remove
this constraint.  In particular, this is true for NSBBN scenarios which cause
a large reduction of the neutron-to-proton ratio
before $\UHe$ is formed, such as ABBN and DBBN.

All nonstandard features,
whose effect on BBN is entirely due to a reduction of
$n/p$ before the nuclei are produced,
have the same effect on BBN constraints on $\DNeff$,
provided a sufficiently large $n/p$ reduction is possible.
One can then consider just this $n/p$ reduction as the nonstandard feature,
and ignore the rest of the nonstandard physics as far as BBN is concerned.

The BBN upper limit to $\DNeff$ is raised, since
this $n/p$ reduction can be cancelled by a corresponding $n/p$ increase due to
the shortening of the timescale caused by a positive $\DNeff$.
This brings $Y_p$ back to the observed value.  A small effect on the other
isotopes remains since the timescale for the nuclear reactions is also
shortened by $\DNeff$.  This can be compensated by an increase in $\eta$.
Thus the larger $\DNeff$ values require somewhat larger values of $\eta$.
The required increase is about $\Delta\et \sim \DNeff/4$.


Allowing for nonstandard BBN removes the upper limit to
additional energy density which comes from nucleosynthesis.
Even when we include
the other cosmological constraints, the limit is raised
from $\DNeff \sim 1$ at least to $\DNeff \sim 4$, or
from $h^2\Omega_x = {\rm few}\times 10^{-6}$ to
$h^2\Omega_x = {\rm few}\times10^{-5}$.
Even values up to $h^2\Omega_x = 8\times10^{-5}$ may be acceptable.
Since, after BBN, gravitational waves interact very weakly
with the surrounding plasma, the upper limit of BBN sets, today, an upper limit
on the critical fraction of energy density stored in gravitational waves
of primordial origin. In the near future various detectors will be
able to reach
  sensitivities, in  $h^2 \Omega_{\rm GW}$, which could be
  $ {\cal O}(10^{-5})$ or even smaller. Hence, the problem will be to understand
whether the obtained signal is primordial or not. The bound on extra
relativistic
species obtained in the context of ABBN
may be then relevant since $h^2 \Omega_{\rm GW}$
cannot certainly be larger than $10^{-4}$ but it can  be larger than
$\sim 5\times10^{-6}$, constraint of
the standard BBN scenario. If a signal compatible with
$h^2 \Omega_{\rm GW} > 5.6 \times 10^{-6}$ will ever be found
(either by two correlated resonant mass detectors
or, most probably, by two correlated Michelson interferometers),
it could still be of primordial origin provided
$h^2 \Omega_{\rm GW} < {\cal O}(10^{-4})$.

\section{Acknowledgments}
We thank the Center for Scientific Computing (Finland) for computational
resources.  On behalf of the
ROG collaboration we  thank V. Fafone and G. Pizzella,
for important discussions. Stimulating  comments
from E. Picasso, G. Gemme, and D. Babusci are also acknowledged.
We thank S.~Pastor, K.~Abazajian, and S.~Hansen for useful comments, S.~Pastor
for additional details concerning the results in Ref.~\cite{Mangano} and
S.~Hansen for the CMB+SNIa contour in Fig.~\ref{fig:allowed}.

\end{document}